\documentclass[twocolumn,floatfix,aps,prb]{revtex4}
\usepackage{graphicx}
\usepackage{amsmath}
\usepackage{amssymb}
\usepackage{bm}
\usepackage{hyperref}

\usepackage{color}

\begin{document}

\title{Quench dynamics of fermion-parity switches in a Josephson junction}
\author{B. Tarasinski}
\affiliation{Instituut-Lorentz, Universiteit Leiden, P.O. Box 9506, 2300 RA Leiden, The Netherlands}
\author{D. Chevallier}
\affiliation{Instituut-Lorentz, Universiteit Leiden, P.O. Box 9506, 2300 RA Leiden, The Netherlands}
\author{Jimmy A. Hutasoit}
\affiliation{Instituut-Lorentz, Universiteit Leiden, P.O. Box 9506, 2300 RA Leiden, The Netherlands}
\author{B. Baxevanis}
\affiliation{Instituut-Lorentz, Universiteit Leiden, P.O. Box 9506, 2300 RA Leiden, The Netherlands}
\author{C. W. J. Beenakker}
\affiliation{Instituut-Lorentz, Universiteit Leiden, P.O. Box 9506, 2300 RA Leiden, The Netherlands}
\date{August 2015}
\begin{abstract}
A Josephson junction may be driven through a transition where the superconducting condensate favors an odd over an even number of electrons. At this switch in the ground-state fermion parity, an Andreev bound state crosses through the Fermi level, producing a zero-mode that can be probed by a point contact to a grounded metal. We calculate the time-dependent charge transfer between superconductor and metal for a linear sweep through the transition. One single quasiparticle is exchanged with charge $Q$ depending on the coupling energies $\gamma_1,\gamma_2$ of the metal to the Majorana operators of the zero-mode. For a single-channel point contact, $Q$ equals the electron charge $e$ in the adiabatic limit of slow driving, while in the opposite quenched limit $Q=2e\sqrt{\gamma_1\gamma_2}/(\gamma_1+\gamma_2)$ varies between $0$ and $e$. This provides a method to produce single charge-neutral quasiparticles on demand.
\end{abstract}
\maketitle

\section{Introduction}
\label{intro}

Superconductors connected by a Josephson junction can freely exchange pairs of electrons, but single-electron transfer is suppressed by the superconducting gap.\cite{Tin04} The tunneling of an unpaired electron into the junction is an incoherent, stochastic source of charge noise in a Cooper pair transistor.\cite{Cla08} In contrast to this undesirable ``quasiparticle poisoning'', a controlled phase-coherent way to exchange single quasiparticles with a superconductor would be a desirable tool, that would complement existing single-electron sources in normal metals and semiconductors.\cite{Fev07,Mah10,Par12,Dub13,Boc13,Boc14}

\begin{figure}[tb]
\centerline{\includegraphics[width=0.9\linewidth]{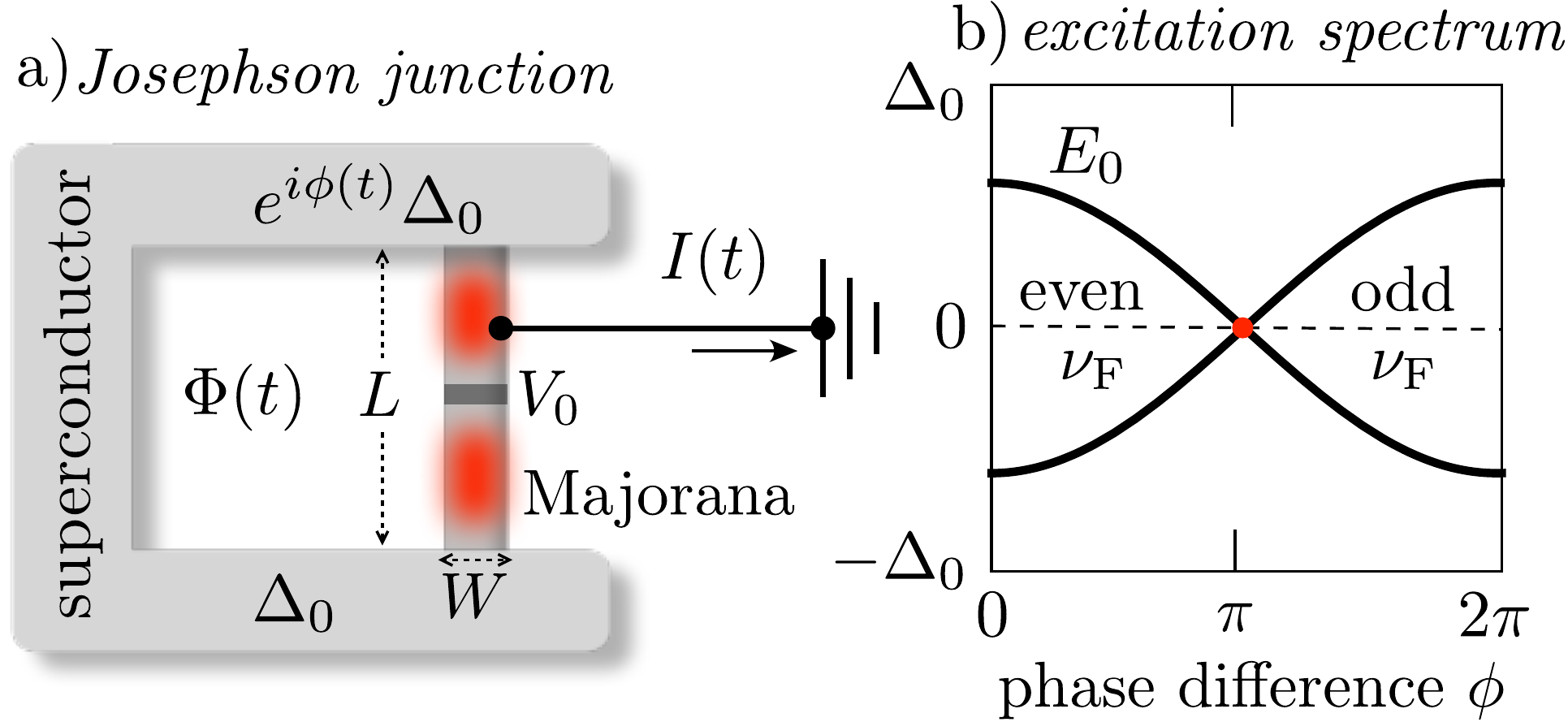}}
\caption{(a) Josephson junction formed by a superconducting ring interrupted by a nanowire. The junction contains two Majorana zero-modes, separated by a tunnel barrier (height $V_0$). A time-dependent flux $\Phi(t)$ through the ring drives the phase $\phi(t)=\Phi(t)\times 2e/\hbar$ through a fermion-parity switch, at which a quasiparticle is injected as a current $I(t)$ into the grounded metal probe. (b) Pair of phase-dependent Andreev levels $\pm E_0(\phi)$ in the closed Josephson junction (uncoupled from the metal). The switch in the ground-state fermion parity $\nu_{\rm F}$ is signaled by a level crossing.
}
\label{fig_layout}
\end{figure}

Here we propose to exploit the phenomenon of a \textit{fermion-parity switch} to transfer, phase coherently and on demand, a single quasiparticle of adjustable charge $Q$ from a Josephson junction to a metal probe (see Fig.\ \ref{fig_layout}a). A fermion-parity switch is a topological phase transition (zero-dimensional class D in the ``ten-fold way'' classification\cite{Alt97,Ryu10}) where the superconducting condensate can lower its ground-state energy by incorporating an unpaired electron and changing the number of electrons in the ground state from $\nu_{\rm F}$ even to $\nu_{\rm F}$ odd,\cite{Bal06} leaving behind as ``defects'' an odd number of quasiparticle excitations above the ground state.

In the quasiparticle excitation spectrum, the switch in the ground-state fermion parity is signaled by the crossing of a pair of bound states (Andreev levels) at $E=0$ (the Fermi level). There may be an even number of switches when the phase difference $\phi$ across the Josephson junction is incremented by $2\pi$ --- if there is an odd number of switches (as in Fig.\ \ref{fig_layout}b) the superconductor is topologically nontrivial. The two lowest Andreev levels $\pm E_0(\phi)$ of a nontrivial Josephson junction have a $\cos(\phi/2)$ phase dependence,\cite{Kit01}
\begin{equation}
E_0(\phi)=\Delta_0\sqrt{T_0}\cos(\phi/2).\label{E0def}
\end{equation}
The superconducting gap is $\Delta_0$ and $T_0\in(0,1)$ is the transmission probability through the junction. For small $T_0$ this describes a pair of bound states at nearly zero energy, consisting of an equal-weight superposition of electron and hole excitations. Such a charge-neutral quasiparticle is called a ``Majorana fermion'' (or Majorana zero-mode) because of the identity of particle and antiparticle. These objects have unusual non-Abelian statistics (see Refs.\ \onlinecite{Wil09,Lei12,Bee13a,Das15} for recent reviews), but here it is only their charge-neutrality that matters.

Fermion-parity switches are actively studied, theoretically\cite{Sau13,Yok13,Bee13,Kes13,Cre14,Hec14} and experimentally,\cite{Lee12,Cha13,Lee14} for the connection to topological superconductivity and Majorana fermions.\cite{Has10,Qi11,Ali13,Bee14} The \textit{dynamics} of the transition is what concerns us here, in particular the quench dynamics, where $\phi(t)$ is driven rapidly through the switch from even to odd ground-state fermion parity.

The geometry of Fig.\ \ref{fig_layout} that we consider is modeled after existing experiments (\textit{e.g.}, Ref.\ \onlinecite{Cha13}), where a mesoscopic Josephson junction is formed by a semiconductor nanowire connecting two arms of a superconducting ring. A time-dependent flux $\Phi(t)$ enclosed by the ring imposes a time dependence on the phase difference $\phi(t)=\Phi(t)\times 2e/\hbar$ across the junction. When the Josephson junction is quenched through a fermion-parity switch there will appear a current pulse $I(t)$ from the superconductor (S) into the metal (N). We seek the quasiparticle content of that pulse. How many quasiparticles are transferred? What is the transferred charge? In particular, we wish to establish the conditions under which a single quasiparticle is transferred with vanishing charge expectation value.

We find that the quench dynamics transfers \textit{one single} quasiparticle from the superconductor to the metal, as a wave packet that is a coherent superposition of electron and hole states near the Fermi level. A nearly charge-neutral equal-weight superposition is produced in a topologically nontrivial superconductor, if the metal probe couples predominantly to one of the two spatially separated Majorana zero-modes. More generally, for two arbitrary coupling constants $\gamma_1,\gamma_2$ we derive that the quantum quench injects a charge
\begin{equation}
Q_{\rm quench}=2e\sqrt{\gamma_1\gamma_2}/(\gamma_1+\gamma_2)\label{Qquench}
\end{equation}
into a single-channel point contact. For a multi-channel point contact the injected charge is reduced further by a factor ${\cal R}$ determined by the peak height $G_{\rm peak}=(4e^2/h)(1-{\cal R}^2)$ of the point contact conductance at resonance.

\section{Microscopic model}
\label{model}

\begin{figure}[tb]
\centerline{\includegraphics[width=0.8\linewidth]{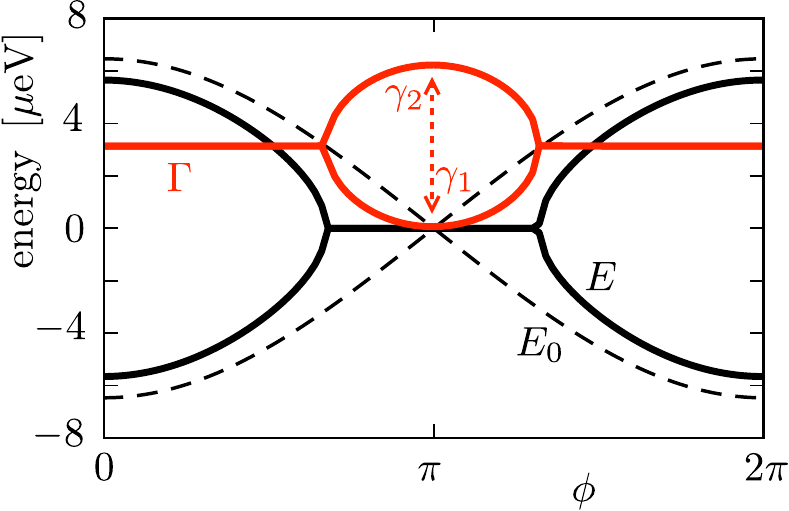}}
\caption{Phase dependence of the complex energies $E_n-i\Gamma_n$ of a pair of quasibound states of the open Josephson junction (solid curves), when the energies $\pm E_0$ of the closed junction (dashed curves) vary through the level crossing of Fig.\ \ref{fig_layout}b. At the fermion parity switch, the inverse lifetimes $\Gamma_n$ reach opposite extremal points $\pi\gamma_n$, $n=1,2$. 
}
\label{fig_poles}
\end{figure}

Before proceeding to the mathematical analysis of the quench dynamics, we explore the relevant physical parameters in a microscopic model\cite{Sta13} for an InSb nanowire (length $L=2.5\,\mu{\rm m}$, width $W=0.25\,\mu{\rm m}$, Fermi energy $E_{\rm F}=1.52\,{\rm meV}$, corresponding to 4 occupied electron subbands), coupled at both ends to a Nb superconductor (induced gap $\Delta_0=0.4\,{\rm meV}$). Spin-rotation symmetry is broken by Rashba spin-orbit coupling (characteristic length $l_{\rm so}=\hbar^2/m_{\rm eff}\alpha_{\rm so}=0.25\,\mu{\rm m}$), and time-reversal symmetry is broken by a magnetic field parallel to the wire (Zeeman energy $V_{\rm Z}=\frac{1}{2}g_{\rm eff}\mu_{\rm B}B=0.6\,{\rm meV}$). For these parameters, the Josephson junction is in the nontrivial regime, with a pair of Majorana zero-modes at the two ends.\cite{Lut10,Ore10} We tune the coupling strength of the Majoranas by means of a tunnel barrier of width $25\,{\rm nm}$ and adjustable height $V_0$ (which might be experimentally realized by means of a gate voltage). The data shown in Fig.\ \ref{fig_poles} is for $V_0=15\,{\rm meV}$. (See App.\ \ref{app_model} for details of the calculation.) 

The Josephson junction is coupled by a point contact to a normal-metal probe, which plays the role of a fermion bath that can exchange quasiparticles with the superconductor. We assume that the charging energy of the junction is much smaller than the Josephson energy, to ensure that the Coulomb blockade of charge transfer is not effective. The Josephson junction is now an open system, with quasibound Andreev states $E_n-i\Gamma_n$ that acquire a finite life time $\hbar/2\Gamma_n$. The evolution of a pair of these states through the fermion-parity switch is shown in Fig.\ \ref{fig_poles}.\cite{kwant} The coupling constants $\gamma_n$ that determine the transferred charge can be read off from
\begin{equation}
\pi\gamma_n={\lim_{\phi\rightarrow\pi}}\;\Gamma_n(\phi).\label{gammanlimphi}
\end{equation}

Particle-hole symmetry requires that the complex energies come in pairs $\pm E- i\Gamma$, symmetrically arranged around the imaginary axis. This constraint produces a bifurcation point (pole transition\cite{Pik12} or exceptional point \cite{San14}) at which the real part is pinned to $E=0$ and the decay rates $\Gamma_1$, $\Gamma_2$ become distinct --- resulting in widely different $\gamma_1$, $\gamma_2$. The unusual extension of the level crossing over a finite interval seen in Fig.\ \ref{fig_poles} is the key distinguishing feature of level crossings in superconducting and non-superconducting systems, and makes the dynamical problem considered here qualitatively different from the familiar Landau-Zener dynamics.\cite{Lan77}

\section{Scattering formulation}
\label{scatteringform}

The exchange of quasiparticles across the NS interface is described by the scattering matrix
\begin{equation}
S(t,t')=\delta(t-t')-2\pi i W^\dagger G(t,t')W.\label{Sdef}
\end{equation}
The coupling matrix $W$ to the fermion bath is assumed to be time-independent. The retarded Green's function $G(t,t')$ satisfies the differential equation\cite{Vav01}
\begin{equation}
\bigl(i\partial/\partial t-H[\phi(t)]+i\pi WW^\dagger\bigr)G(t,t')=\delta(t-t'),\label{Gdef}
\end{equation}
where $H(\phi)$ is the Bogoliubov-De Gennes Hamiltonian of the Josephson junction at a fixed value $\phi$ of the superconducting phase difference. (We have set $\hbar\equiv 1$ for ease of notation.) Fourier transform to the energy domain is defined by
\begin{equation}
S(E,E')=\int_{-\infty}^\infty dt\,\int_{-\infty}^\infty dt'\,e^{iEt-iE't'}S(t,t').\label{SFourier}
\end{equation}

In a stationary situation, with a time-independent Hamiltonian $H$, the scattering matrix is diagonal in energy, $S(E,E')=2\pi\delta(E-E')\,S_0(E)$, with $S_0$ given by the Mahaux-Weidenm\"{u}ller formula,\cite{Mah69}
\begin{equation}
\begin{split}
&S_0(E)=1-2\pi i W^\dagger(E-H_{\rm eff})^{-1}W,\\
&H_{\rm eff}=H-i\pi WW^\dagger.
\end{split}\label{S0def}
\end{equation}
The formulation of this dynamical problem in an open system in terms of an effective non-Hermitian Hamiltonian $H_{\rm eff}$ goes back to the early days of nuclear scattering theory.\cite{Liv57,Fes58}

For a minimal description, we take a pair of Andreev levels in the Josephson junction coupled to a pair of electron-hole modes in a single-channel metal probe. (The multi-channel case is addressed in Sec.\ \ref{multichannelprobe}.) Both $H$ and $W$ are now $2\times 2$ matrices. Particle-hole symmetry requires that
\begin{equation}
H=-\sigma_x H^\ast\sigma_x,\;\;W=\sigma_x W^\ast \sigma_x.\label{phsymmetry}
\end{equation}
(The Pauli matrix $\sigma_x$ interchanges electron and hole indices.) Particle-hole symmetry is the only symmetry constraint we impose on the system (symmetry class D), assuming that time-reversal symmetry and spin-rotation symmetry are both broken by magnetic field and spin-orbit coupling in the nanowire.

Using also that $H=H^\dagger$, we have the general form
\begin{equation}
H=E_0\sigma_z,\;\;W=e^{i\alpha'\sigma_z}\Lambda e^{i\alpha\sigma_z},\;\;\Lambda=\begin{pmatrix}
\lambda_+&\lambda_-\\
\lambda_-&\lambda_+
\end{pmatrix},\label{HWgeneral}
\end{equation}
with real coefficients $\alpha,\alpha'$, $\lambda_\pm$. The eigenvalues $\gamma_1,\gamma_2\geq 0$ of the coupling matrix product $WW^\dagger$ are given by
\begin{equation}
\gamma_1=(\lambda_+ +\lambda_-)^2,\;\; \gamma_2=(\lambda_+ - \lambda_-)^2.\label{gammawrelation}
\end{equation}

\begin{figure}[tb]
\centerline{\includegraphics[width=0.8\linewidth]{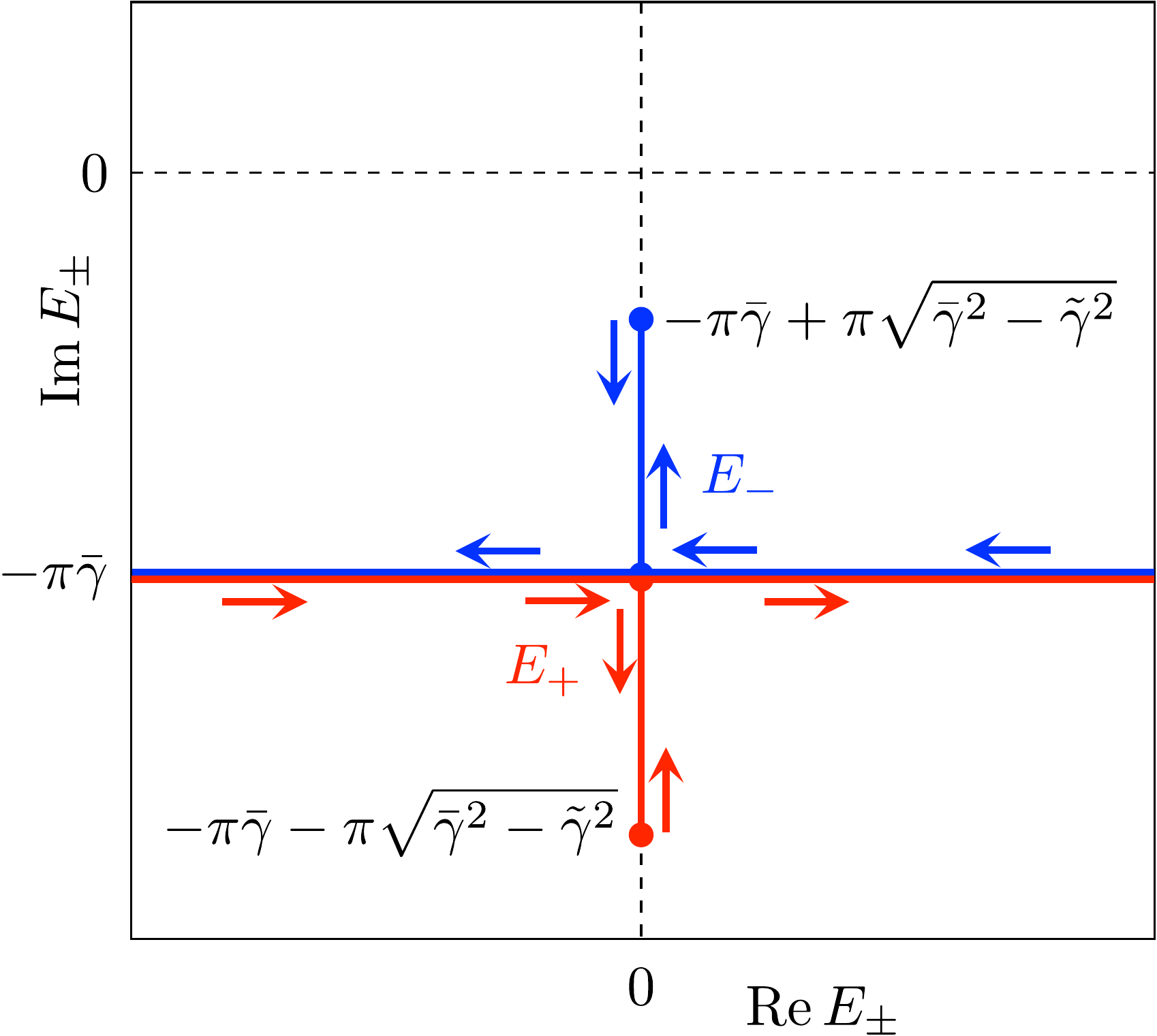}}
\caption{Evolution of the complex eigenvalues $E_\pm$ of the effective Hamiltonian \eqref{S0def} of the open Josephson junction (coupled to a metal probe), when the real eigenvalues $\pm E_0$ of the closed junction vary through a level crossing. At the fermion parity switch, $E_0=0$  and $E_\pm$ reach opposite extremal points on the imaginary axis.
}
\label{fig_bifurcation}
\end{figure}

The eigenvalues of $H_{\rm eff}$ (representing the poles of $S_0$ in the complex energy plane) are given by
\begin{equation}
E_\pm=-i\pi\bar{\gamma}\pm E_0\sqrt{1+(\pi\tilde{\gamma}/E_0)^2-(\pi\bar{\gamma}/E_0)^2},\label{Epmdef}
\end{equation}
in terms of the arithmetic and geometric mean
\begin{equation}
\bar{\gamma}=\tfrac{1}{2}(\gamma_1+\gamma_2),\;\;\tilde{\gamma}=\sqrt{\gamma_1\gamma_2}.\label{bartildegamma}
\end{equation}
The evolution of $E_\pm$ through the fermion-parity switch is shown in Fig.\ \ref{fig_bifurcation}. The relation $E_+=-E_-^\ast$ required by particle-hole symmetry produces a bifurcation point at which the two quasibound states acquire distinct decay rates,\cite{Pik12,San14} see also Fig.\ \ref{fig_poles}.

The time dependent phase difference $\phi(t)$ across the Josephson junction shakes up the fermion bath in the normal metal. We assume zero temperature, so that the unperturbed Fermi sea is the vacuum state $|0\rangle$ for excitations: $a(E)|0\rangle=0$ for $E>0$, with $a=(a_1,a_2)$ the two-component Nambu spinor of annihilation operators for Bogoliubov quasiparticles. The fermion-parity switch produces a superposition 
\begin{equation}
|\Psi\rangle=\zeta_0|0\rangle+\textstyle{\sum_{p=1}^\infty}|\Psi_p\rangle\label{zeta0def}
\end{equation} 
of the vacuum state with $p$-particle excited states
\begin{equation}
|\Psi_p\rangle=\left[\sum_{E>0}\sum_{E'<0} a^\dagger(E)S(E,E')a(E')\right]^p|0\rangle.\label{Psidef}
\end{equation}
(The sum $\sum_E$ is evaluated as $(2\pi)^{-1} \int dE$.) The weight $\zeta_0$ of the unperturbed Fermi sea follows from the normalization $\langle\Psi|\Psi\rangle=1$.

\section{Linear sweep through the fermion-parity switch}
\label{exactsolution}

We now proceed to a complete solution of the dynamics of the fermion-parity switch, to derive the result \eqref{Qquench} for the charge of the transferred quasiparticle. The non-superconducting counterpart to this problem was studied by Keeling, Shytov, and Levitov.\cite{Kee08} Their analysis provided much guidance and inspiration for what follows.

We calculate the scattering matrix for a linear sweep through the fermion parity switch: $E_0[\phi(t)]=\gamma_0^2 t$. Referring to Eq.\ \eqref{E0def}, this linear approximation of the spectrum is justified for rapidities $\gamma_0^2\ll\sqrt{T_0}\Delta_0\bar{\gamma}$. In the energy domain, Eqs.\ \eqref{Sdef} and \eqref{Gdef} then take the form
\begin{align}
&S(E,E')=2\pi\delta(E-E')-2\pi i e^{-i\alpha\sigma_z}\Lambda G(E,E')\Lambda e^{i\alpha\sigma_z},\nonumber\\
&\bigl(i \gamma_0^2\sigma_z\partial/\partial E+E +i\pi \Lambda^2 \bigr)G(E,E')=2\pi\delta(E-E').\label{GdefM}
\end{align}

The solution for the Green's function factorizes,
\begin{align}
&G(E,E')=\frac{2\pi}{i\gamma_0^2}X(E)\Theta(E-E')\sigma_z X^{-1}(E')\sigma_z,\label{Gfactorized}\\
&\Theta(E-E')=\begin{pmatrix}
\theta(E-E')&\!\!\!\!0\\
0&\!\!\!\!\theta(E'-E)
\end{pmatrix}.\label{Thetadef}
\end{align}
Here $\theta(E)$ is the unit step function and the matrix $X(E)$ solves the homogeneous equation\cite{note1}
\begin{equation}
\bigl(i \gamma_0^2\sigma_z\partial/\partial E+E +i\pi \Lambda^2 \bigr)X(E)=0.\label{Xdef}
\end{equation}
Because of particle-hole symmetry, $X$ has two rather than four independent elements,
\begin{equation}
X(E)=\begin{pmatrix}
u(E)&v^\ast(-E)\\
v(E)&u^\ast(-E)
\end{pmatrix},\label{Xuvdef}
\end{equation}
determined by
\begin{align}
&\gamma_0^2 u'' + (\varepsilon^2+\delta^2-i)u=0,\;\;
\delta v=i\varepsilon u-\gamma_0 u',
\label{eq:second-order-eq-for-u}\\
&\varepsilon=(E+i\pi\bar{\gamma})/\gamma_0,\;\;\delta=\tfrac{1}{2}\pi(\gamma_1-\gamma_2)/\gamma_0.\label{epsdeltadef}
\end{align}

The retarded Green's function is specified by $G\rightarrow 0$ in the limits $E\rightarrow+\infty$ or $E\rightarrow-\infty$. The factor $\Theta$ in Eq.\ \eqref{Gfactorized} ensures that this two-sided decay follows from the one-sided decay $u,v\rightarrow 0$ for $E\rightarrow +\infty$. With this condition the solution of Eq.\ \eqref{eq:second-order-eq-for-u} reads\cite{note2}
\begin{equation}
\begin{split}
&u(E) = e^{i\varepsilon^2/2}\,U( -\tfrac{1}{4}i \delta^2,\tfrac{1}{2}; -i\varepsilon^2  ), \\
&v(E) = -\tfrac{1}{2}\delta e^{i\pi/4} \,e^{i\varepsilon^2/2}\,U( \tfrac{1}{2}-\tfrac{1}{4}i \delta^2,\tfrac{1}{2}; -i\varepsilon^2  ),
\end{split}
\label{eq:solution-in-terms-of-u}
\end{equation}
where $U$ is the confluent hypergeometric function of the second kind.\cite{Abr72,Leb72} The determinant of $X$ is particularly simple (see App.\ \ref{app_details})
\begin{equation}
{\rm Det}\,X=\exp(-\pi\delta^2/4),\label{DetXresult}
\end{equation}
independent of energy.

The scattering matrix \eqref{GdefM} results as the dyadic product of two vectors,
\begin{align}
&S_{nm}(E,E')|_{E>E'}=-\psi_n(E)\psi^\ast_m(-E'),\label{Suv}\\
&\psi(E)=(2\pi/\gamma_0)e^{\pi\delta^2/8}
e^{-i\alpha\sigma_z}\Lambda
\begin{pmatrix}
u(E)\\
v(E)
\end{pmatrix}.\label{psidef}
\end{align}
Substitution into Eq.\ \eqref{Psidef} gives $|\Psi_p\rangle=0$ for $p\geq 2$ because of the anticommutation of the creation operators, so that only a single-particle excitation remains,\cite{note3}
\begin{equation}
|\Psi_1\rangle=-\sum_{E>0}\sum_{E'<0}[\psi(E)a^\dagger(E)][\psi^\ast(-E')a(E')]|0\rangle.
\label{PsiX}
\end{equation} 
This absence of multi-particle excitations is a generic feature of rank-one scattering matrices.\cite{Kee08,Kee06}

The normalization $\sum_{E>0}|\psi(E)|^2=1$ can be derived directly from Eq.\ \eqref{Xdef}. (See App.\ \ref{app_details}.) This implies that $\langle\Psi_1|\Psi_1\rangle=1$, hence there is no contribution from the vacuum state [$\zeta_0=0$ in Eq.\ \eqref{zeta0def}]. Corrections of order $|e^{i\varepsilon^2}|=\exp(-2\pi E\bar{\gamma}/\gamma_0^2)$ to the normalization appear because of the finite band width $E\lesssim \sqrt{T_0}\Delta_0$. Since we have assumed $\gamma_0^2\ll\sqrt{T_0}\Delta_0\bar{\gamma}$ we can ascertain that the sweep through the fermion-parity switch will fail to produce a quasiparticle with exponentially small probability.

The Josephson junction thus injects a \textit{single} Bogoliubov quasiparticle into the metal probe, in a pure state with wave function $\psi$ given by Eq.\ \eqref{psidef}. The transfer of this quasiparticle is observable as an electrical current pulse, with expectation value
\begin{equation}
I(t)=e\int_0^\infty\frac{dE}{2\pi}\int_0^\infty\frac{dE'}{2\pi}\,e^{i(E'-E)t}\psi^\ast(E')\sigma_z\psi(E).\label{Itdef}
\end{equation}
The expectation value of the total transferred charge $Q=\int_{-\infty}^\infty I(t)dt$ is given by
\begin{equation}
Q=\frac{2\pi e}{\gamma_0^2}(\lambda_+^2-\lambda_-^2)\,e^{\pi\delta^2/4}\int_{0}^{\infty}dE\,\bigl(|u(E)|^2-|v(E)|^2\bigr).\label{Qintegral}
\end{equation}
For definiteness we take $\lambda_+^2\geq \lambda_-^2$ in what follows (otherwise the sign of currents and charges should be inverted).

\begin{figure}[tb]
\centerline{\includegraphics[width=0.8\linewidth]{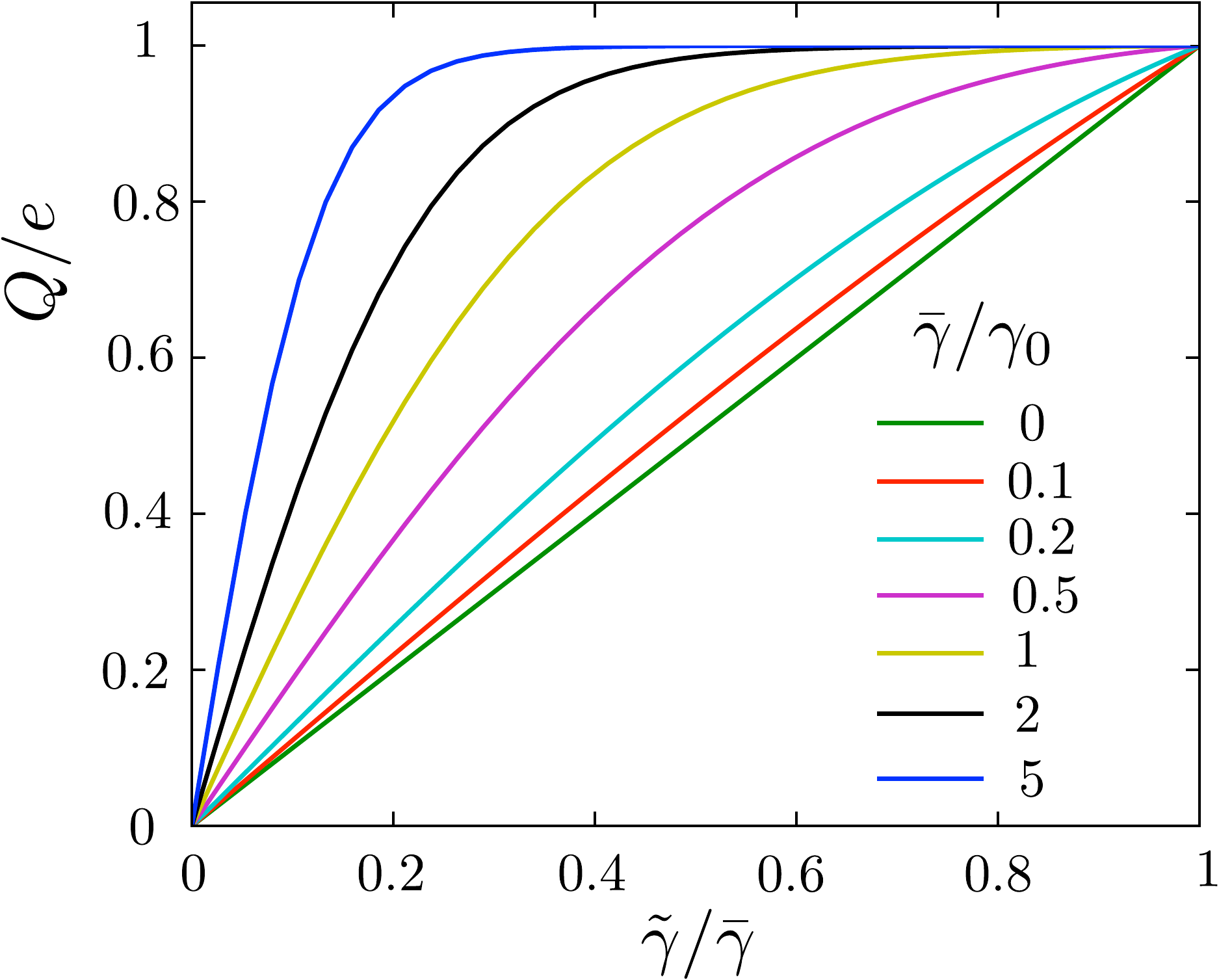}}
\caption{Expectation value of the charge of the quasiparticle transferred between the superconductor and a single-channel metal probe, following a fermion-parity switch with rapidity $\gamma_0$. The charge $Q$ is given as a function of the ratio $\tilde{\gamma}/\bar{\gamma}$ of the geometric and arithmetic mean of the coupling energies to the two Majorana operators involved in the transition. The curves are calculated numerically from Eq.\ \eqref{Qintegral}. The quenched and adiabatic limits are given by Eqs.\ \eqref{Qquenchresult} and \eqref{Qadresult}.
}
\label{fig_charge}
\end{figure}

\section{Transferred charge}
\label{transcharge}

\subsection{Single-channel probe}
\label{singlechannelprobe}

A single quasiparticle passes through the NS interface irrespective of the rapidity $\gamma_0$, but the transferred charge differs. Fig.\ \ref{fig_charge} shows results from a numerical evaluation of Eq.\ \eqref{Qintegral}. Analytical results can be obtained in the quenched limit $\gamma_0\gg \gamma_1,\gamma_2$ of a fast fermion-parity switch and in the opposite adiabatic limit $\gamma_0\ll\gamma_1,\gamma_2$ of a slow switch. 

In the quenched limit we set $\delta\rightarrow 0$ and since $U(0,\tfrac{1}{2};-i\varepsilon^2)=1$ we have $u\rightarrow\exp(i\varepsilon^2/2)$, $v\rightarrow 0$. The current and transferred charge evaluate to
\begin{equation}
I_{\rm quench}(t)=2\pi e\tilde{\gamma} \exp(-2\pi\bar{\gamma}t)\theta(t),\;\;Q_{\rm quench}=e\tilde{\gamma}/\bar{\gamma}.\label{Qquenchresult}
\end{equation}
This is the result \eqref{Qquench} announced in the introduction.

The adiabatic limit may be obtained, with some effort, from the Fourier transform \eqref{Itdef} in saddle-point approximation, or more easily by starting directly from the general scattering formula\cite{But94,Bro98,Avr00,Bla02}
\begin{equation}
I_{\rm adiabatic}(t)=\frac{ie}{4\pi}\,{\rm Tr}\,S_{\rm F}^\dagger(0,t)\sigma_z \frac{\partial}{\partial t}S_{\rm F}(0,t).\label{IBPT}
\end{equation}
(A selfcontained derivation of this formula is given in App.\ \ref{app_adiabatic}.) The adiabatic charge transfer is described by the ``frozen'' scattering matrix
\begin{equation}
S_{\rm F}(E,t)=S_0(E)|_{\phi\equiv\phi(t)},\label{Sfrozen}
\end{equation}
with $S_0$ from Eq.\ \eqref{S0def} evaluated for a fixed value $\phi(t)$ of the phase across the Josephson junction. The result is
\begin{equation}
I_{\rm adiabatic}(t)=\frac{e\sqrt{\gamma_1\gamma_2}}{\pi^2\gamma_1\gamma_2/\gamma_0^2+\gamma_0^2 t^2},\;\;
Q_{\rm adiabatic}=e.\label{Qadresult}
\end{equation}

The exponential versus Lorentzian current profiles \eqref{Qquenchresult} and \eqref{Qadresult} have the same form as in the non-superconducting problem of Ref.\ \onlinecite{Kee08}, but there the transferred quasiparticle was an electron of charge $e$. Here what is transferred is a Bogoliubov quasiparticle, which is not in an eigenstate of charge. In the quenched limit $Q$ can vary between $0$ and $e$, depending on the ratio of the geometric and arithmetic mean of the two coupling energies $\gamma_1$, $\gamma_2$ of the metal probe to the Majorana operators of the zero-mode. A nearly charge-neutral quasiparticle is transferred if $\gamma_1\ll\gamma_2$, when $Q=2e\sqrt{\gamma_1/\gamma_2}$ in the quenched limit.

\subsection{Multi-channel probe}
\label{multichannelprobe}

So far we have assumed that the metal probe supports a single electron-hole channel. More generally, the coupling between the superconductor and the metal would involve $N$ electron-hole channels, where $N$ would include both orbital and spin degrees of freedom. This multi-channel generalization is worked out in App.\ \ref{app_multi}. A single quasiparticle is injected, as before, with a reduced charge $Q_{N}={\cal R}Q_1$. The reduction factor ${\cal R}\in[0,1]$ is independent of the rapidity $\gamma_0$. It is determined entirely by the point contact conductance, which at the fermion parity switch has a resonant peak of height
\begin{equation}
G_{\rm peak}=\frac{4e^2}{h}(1-{\cal R}^2).\label{Gpeakdef}
\end{equation}

\section{Conclusion}
\label{conclude}

In conclusion, we have investigated the phase-coherent, deterministic counterpart of incoherent, stochastic quasiparticle poisoning: A fermion-parity switch in a Josephson junction transfers a single quasiparticle into a metal contact, on demand and in a pure state. The quasiparticle is a coherent superposition of electron and hole, with a charge expectation value that can be adjusted between 0 and $e$. A nearly charge-neutral quasiparticle is produced in the quenched limit of a fast parity switch, if the metal couples predominantly to a single Majorana operator in the Josephson junction. This device could be used for superconducting analogues of single-electron collision experiments,\cite{Fev07,Mah10,Par12,Dub13,Boc13,Boc14} such as the Hanbury-Brown--Twiss or Hong-Ou-Mandel interferometer for Majorana fermions.\cite{Bee14b,Fer15}

Experimentally, one can determine the value of $Q$ by sweeping up and down through the fermion-parity switch and measuring the shot noise power $P_{\rm shot}$. In each period $\tau$ a charge $\{0,+e,-e\}$ is transferred with probability $\{1-2p(1-p),p(1-p),p(1-p)\}$, where $Q/e=|1-2p|$ is the average charge transferred during a sweep up or down. The full distribution of the transferred charge is trinomial. The first moment vanishes and the second moment is given by
\begin{equation}
P_{\rm shot}=2p(1-p)(e^2/\tau)=\tfrac{1}{2}\tau^{-1}(e^2-Q^2).\label{Pshot}
\end{equation}

Referring to the model calculation of Fig.\ \ref{fig_poles}, a band width of ${\sqrt T_0}\Delta_0\simeq 10\,{\rm GHz}$ at a driving frequency of $1/\tau\simeq 0.1\,{\rm GHz}$ would imply a rapidity $\gamma_0\simeq 1\,{\rm GHz}$ (so that $\gamma_0^2\tau\simeq{\sqrt T_0}\Delta_0$). The escape rate $\bar{\gamma}$ could then vary between, say, $0.2\,{\rm GHz}$ and $2\,{\rm GHz}$ to vary between the adiabatic and the quenched regime. These frequencies should all lie above the decoherence rate of the Bogoliubov quasiparticle due to charge noise, which could be below 1~MHz.\cite{Sch12}

An alternative way to measure the transferred charge is to apply a voltage $V$ between the two superconductors. The phase will then advance with constant rate $d\phi/dt=2eV/\hbar$, producing a current $I=Q\times 2eV/h$ (assuming a single level crossing in a $2\pi$ phase interval).

\acknowledgments

We acknowledge discussions with R. Aguado, P. W. Brouwer, S. Mi, and P. San-Jose. This research was supported by the Foundation for Fundamental Research on Matter (FOM), the Netherlands Organization for Scientific Research (NWO/OCW), and an ERC Synergy Grant.

\appendix

\section{Model Hamiltonian}
\label{app_model}

\begin{figure}[tb]
\centerline{\includegraphics[width=1\linewidth]{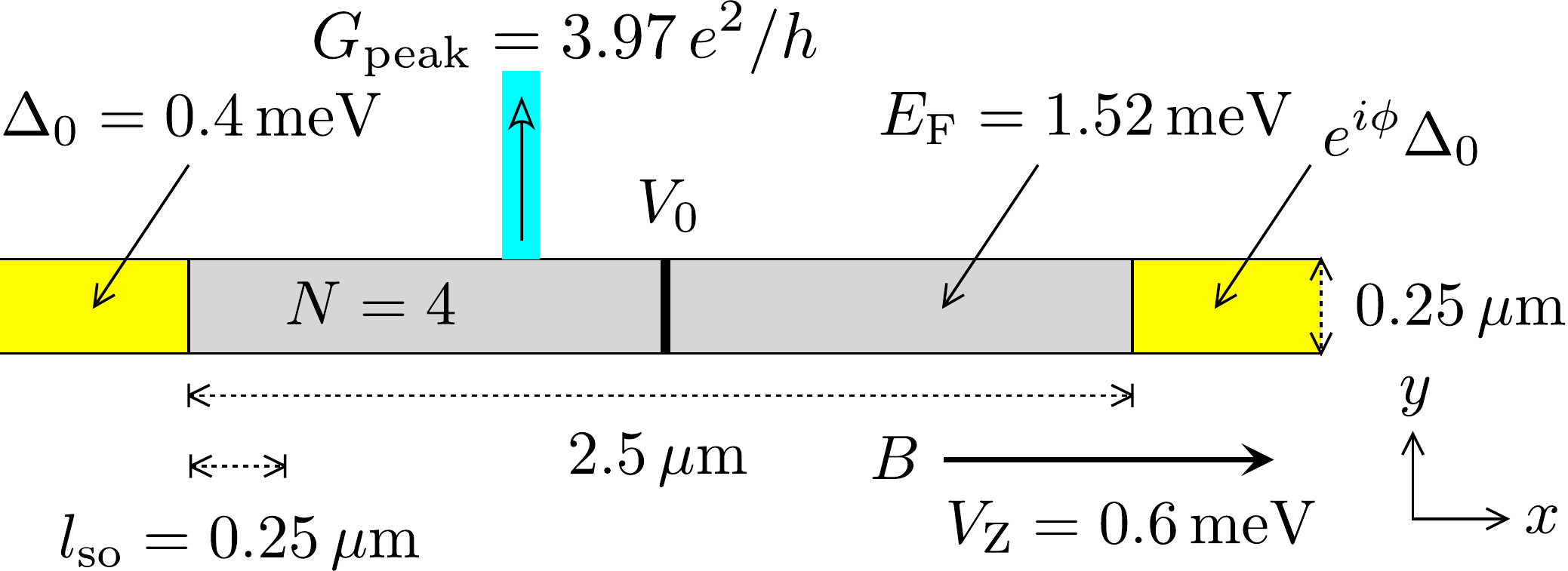}}
\caption{Nanowire Josephson junction modeled by the Hamiltonian \eqref{TBH}, discretized on a square lattice (lattice constant $a=25\,{\rm nm}$). The InSb nanowire is grey, with a tunnel barrier (width $25\,{\rm nm}$) in black, the superconducting contacts are yellow, the normal-metal probe (width $100\,{\rm nm}$) is blue. There are 4 electron subbands in the nanowire and 8 in the probe, counting spin. The peak conductance at the fermion-parity switch is indicated.
}
\label{fig_tightbinding}
\end{figure}

The model Hamiltonian for the nanowire Josephson junction of Fig.\ \ref{fig_tightbinding} has the Bogoliubov-De Gennes form
\begin{subequations}
\label{TBH}
\begin{align}
H={}&\begin{pmatrix}
H_0(\mathbf{p})  & \Delta \\
\Delta^{\ast} & - \sigma_y H_0^{\ast} (-\mathbf{p}) \sigma_y
\end{pmatrix},\\
H_0 = {}& \frac{\mathbf{p}^2}{2m_{\rm eff}} - E_{\rm F} + \frac{\alpha_{\rm so}}{\hbar} (\sigma_x p_y - \sigma_y p_x) + \tfrac{1}{2} g_{\rm eff}\mu_{\rm B} B \sigma_x \nonumber\\
& + V_0~[\Theta(x-W_{\rm B}/2) - \Theta(x-W_{\rm B}/2)] .
\end{align}
\end{subequations}
Electrons and holes are coupled by the induced \textit{s}-wave pair potential $\Delta$ at the superconducting contacts, with a phase difference $\phi$. The single-particle Hamiltonian $H_0$ contains Rashba spin-orbit coupling and the Zeeman energy of a magnetic field parallel to the nanowire. A potential barrier of strength $V_0$ and width $W_{\rm B}$ is located at the center of the junction.

The Hamiltonian $H$ is discretized on a square lattice, to obtain a tight-binding model.\cite{kwant} For the parameters indicated in the figure, the Josephson junction is in the nontrivial regime,\cite{Lut10,Ore10} with a pair of Majorana zero-modes at the normal-superconducting (NS) interface, weakly coupled via the potential barrier. A normal-metal lead is attached perpendicular to the nanowire, coupling predominantly to one of the two zero-modes.

To obtain the complex energies of the quasibound states, the imaginary part of the lead self-energy is added to the tight-binding Hamiltonian of the junction. Diagonalization of this non-Hermitian Hamiltonian yields the complex eigenvalues $ E_n(\phi) - i\Gamma_n(\phi)$ plotted in Fig.\ \ref{fig_poles}.

\section{Details of the calculation of the Green's function}
\label{app_details}

\subsection{Evaluation of the determinant}

Since the expression \eqref{Gfactorized} for the Green's function contains both the matrix $X(E)$ and its inverse, we need to evaluate the determinant of this $2\times 2$ matrix. As a first step we will show that ${\rm Det}\,X$ is energy independent. This can be done directly from the differential equation \eqref{Xdef} for $X$.

We write the determinant in the form
\begin{equation}
{\rm Det}\,X(E)=\begin{pmatrix}
u^{*}(-E)\\
v^{*}(-E)
\end{pmatrix}^{\rm T}
\sigma_z
\begin{pmatrix}
u(E)\\
v(E)
\end{pmatrix},\label{determinantX}
\end{equation}
and take the derivative with respect to $E$. The functions $u,v$ solve
\begin{equation}
\bigl(i \gamma_0^2\sigma_z d/dE+E +i\pi \Lambda^2 \bigr)\begin{pmatrix}
u\\
v
\end{pmatrix}=0.\label{Xdefapp}
\end{equation}
This allows us to express the derivatives
\begin{align}
\frac{d}{dE}\begin{pmatrix}
u(E)\\
v(E)
\end{pmatrix}={}&\frac{i}{\gamma_0^2}\sigma_z(E+i\pi\Lambda^2)\begin{pmatrix}
u(E)\\
v(E)
\end{pmatrix},\\
\frac{d}{dE}\begin{pmatrix}
u^\ast(-E)\\
v^\ast(-E)
\end{pmatrix}={}&-\frac{i}{\gamma_0^2}\sigma_z(E+i\pi{\Lambda^\ast}^2)\begin{pmatrix}
u^\ast(-E)\\
v^\ast(-E)
\end{pmatrix}.
\end{align}
Since $\Lambda$ is a real and symmetric matrix, it follows that 
\begin{align}
\frac{d}{dE}\,{\rm Det}\,X={}&\frac{i}{\gamma_0^2}\begin{pmatrix}
u^{*}(-E)\\
v^{*}(-E)
\end{pmatrix}^{\rm T}
[(E+i\pi\Lambda^2)\nonumber\\
&-(E+i\pi{\Lambda^\ast}^2)^{\rm T}]\begin{pmatrix}
u(E)\\
v(E)
\end{pmatrix}\nonumber\\
={}&0,
\end{align}
so ${\rm Det}\,X$ is independent of $E$.

From Eq.\ \eqref{eq:solution-in-terms-of-u} we have an explicit expression for the determinant of $X$:
\begin{align}
&{\rm Det}\,X=U( -\tfrac{1}{4}i \delta^2,\tfrac{1}{2}; -i\varepsilon^2  )U(\tfrac{1}{4}i \delta^2,\tfrac{1}{2}; i\varepsilon^2  )\nonumber\\
&\quad-\tfrac{1}{4}\delta^2 U( \tfrac{1}{2}-\tfrac{1}{4}i \delta^2,\tfrac{1}{2}; -i\varepsilon^2  )U( \tfrac{1}{2}+\tfrac{1}{4}i \delta^2,\tfrac{1}{2}; i\varepsilon^2  ).\label{DetXU}
\end{align}
This is an analytic function of $\varepsilon=(E+i\pi\bar{\gamma})/\gamma_0$, which is independent of $E$ and hence independent of $\varepsilon$. At $\varepsilon=0$ we may evaluate it by means of the identities\cite{Abr72}
\begin{align}
&U(a,\tfrac{1}{2},0)=\frac{\sqrt{\pi}}{\Gamma(\tfrac{1}{2}+a)},\\
&\Gamma(\tfrac{1}{2}+ia)\Gamma(\tfrac{1}{2}-ia)=\frac{\pi}{\cosh\pi a},\nonumber\\
&\Gamma(1+ia)\Gamma(1-ia)=\frac{\pi a}{\sinh\pi a}.
\end{align}
Substitution into Eq.\ \eqref{DetXU} at $\varepsilon=0$ gives
\begin{equation}
{\rm Det}\,X=\exp(-\pi\delta^2/4),
\end{equation}
as in Eq.\ \eqref{DetXresult}.

\subsection{Normalization of the excited state}

We wish to demonstrate that the wave function \eqref{psidef} of the single-particle excited state is normalized to unity. For that purpose we need to evaluate the integral
\begin{equation}
{\cal N}\equiv\langle \psi|\psi\rangle =\int_{0}^{\infty}\frac{2\pi dE}{\gamma_0^2\,{\rm Det}\,X}
\begin{pmatrix}
u^{*}(E)\\
v^{*}(E)
\end{pmatrix}^{\rm T}
\Lambda^2
\begin{pmatrix}
u(E)\\
v(E)
\end{pmatrix}.\label{normalization}
\end{equation}

We again use the fact that $u,v$ solve Eq.\ \eqref{Xdefapp}. Substitution into Eq.\ \eqref{normalization} gives (denoting $u'=du/dE$)
\begin{align}
{\cal N}={}&\frac{-2}{{\rm Det}\,X}\int_{0}^{\infty}dE\,\bigl[u^{*}u'-v^{*}v'-iE\gamma_0^{-2}(uu^{*}+vv^{*})\big]\nonumber\\
={}&\frac{2}{{\rm Det}\,X}\bigl(|u(0)|^{2}-|v(0)|^{2}\bigr)\nonumber\\
&\!\!\!\!\!+\frac{2}{{\rm Det}\,X}\int_{0}^{\infty}dE\,\bigl[u{u^{*}}'-v{v^{*}}'+iE\gamma_0^{-2}(uu^{*}+vv^{*})\bigr]\nonumber\\
={}&2-{\cal N}^{*},
\end{align}
and because ${\cal N}$ is real, we indeed have ${\cal N}=1$. Notice that $\langle\psi|\psi\rangle=1$ also implies $\langle\Psi_1|\Psi_1\rangle=1$ in Eq.\ \eqref{PsiX}.

\section{Scattering formula for the charge transfer in the adiabatic regime}
\label{app_adiabatic}

The current passing through the NS interface in the adiabatic regime $\gamma_0\ll\gamma_1,\gamma_2$ of a slow fermion-parity switch can be evaluated most easily from the scattering formula \eqref{IBPT}, which is the analogue for Bogoliubov quasiparticles of a well-known formula for normal electrons.\cite{But94,Bro98,Avr00,Bla02} For completeness we give a derivation of Eq.\ \eqref{IBPT}.

One subtlety in this derivation is that Fourier transforms of quasiparticle annihilation operators $a(E)$ to the time domain need to include both positive and negative energies in order to produce a complete basis set. This results in a double counting of the quasiparticle excitations, because of the relation $a(-E)=\sigma_x a^\dagger(E)$. To correct for the double counting we include a factor $1/2$ in the definition of the current operator,\cite{Bee14b}
\begin{equation}
\begin{split}
&{\cal I}(t)=\tfrac{1}{2}ea_{\rm out}^\dagger(t)\sigma_z a_{\rm out}(t),\\
&a_{\rm out}(t)=\int_{-\infty}^\infty\frac{dE}{2\pi}\,e^{-iEt}a_{\rm out}(E).
\end{split}\label{Ioperatort}
\end{equation}

The outgoing and incoming operators are related by the scattering matrix,
\begin{equation}
a_{\rm out}(E)=\int_{-\infty}^\infty \frac{dE'}{2\pi} S(E,E')a_{\rm in}(E'),\label{aoutain}
\end{equation}
which satisfies the unitarity condition
\begin{equation}
\begin{split}
\int_{-\infty}^{\infty}\frac{dE'}{2\pi}\,\sum_{n'}S_{nn'}(E_1,E')S^\ast_{mn'}(E_2,E')\\
=2\pi\delta_{nm}\delta(E_1-E_2).
\end{split}
\label{Sunitary}
\end{equation}

The incoming operators have the equilibrium expectation value
\begin{equation}
\langle a_{n}^\dagger(E)a_{m}(E')\rangle=2\pi\delta(E-E')\delta_{nm}f(E),\label{aexpectation}
\end{equation}
with $f(E)=(1+e^{E/kT})^{-1}$ the Fermi function at temperature $T$. We seek the current expectation value $I(t)\equiv\langle{\cal I}(t)\rangle$, given by
\begin{align}
I(t)={}&\tfrac{1}{2}e\int_{-\infty}^\infty\frac{dE}{2\pi}\int_{-\infty}^\infty\frac{dE'}{2\pi}\int_{-\infty}^\infty\frac{d\omega}{2\pi}\,e^{i\omega t}\nonumber\\
&\times f(E')\,{\rm Tr}\,S^\dagger(E+\omega,E')\sigma_z S(E,E').\label{Itaverage0}
\end{align}

Because of the unitarity condition \eqref{Sunitary}, the integral over $E'$ without the factor $f(E')$ vanishes,
\begin{align}
\int_{-\infty}^\infty \frac{dE'}{2\pi}\,{\rm Tr}\,S^\dagger(E+\omega,E')\sigma_z S(E,E')&=2\pi\delta(\omega)\,{\rm Tr}\,\sigma_z\nonumber\\
&=0.\label{vanishingtrace}
\end{align}
We may therefore equivalently write
\begin{align}
I(t)={}&\tfrac{1}{2}e\int_{-\infty}^\infty\frac{dE}{2\pi}\int_{-\infty}^\infty\frac{dE'}{2\pi}\int_{-\infty}^\infty\frac{d\omega}{2\pi}\,e^{i\omega t}\nonumber\\
&\times [f(E')-f(E)]\, {\rm Tr}\,S^\dagger(E+\omega,E')\sigma_z S(E,E').\label{Itaverage1}
\end{align}

It is convenient to introduce the Wigner transform
\begin{equation}
S_{\rm W}(E,t)=\int_{-\infty}^\infty\frac{dE'}{2\pi}\,e^{-iE't}S(E+\tfrac{1}{2}E',E-\tfrac{1}{2}E'),\label{SWdef0}
\end{equation}
because it becomes the frozen scattering matrix $S_{\rm F}(E,t)$ from Eq.\ \eqref{Sfrozen} in the adiabatic limit.\cite{Vav01} More precisely,
\begin{equation}
S_{\rm W}(E+\delta E,t)=S_{\rm F}(E,t)+{\cal O}(\gamma_0/E_{\rm c})+{\cal O}(\delta E/E_{\rm c}),
\label{SWSFexpansion}
\end{equation}
with $E_{\rm c}=\min(\gamma_1,\gamma_2)$ the width of the quasi-bound state.

Fourier transformation of the time variable gives
\begin{equation}
S_{\rm W}(E,\omega)=\int_{-\infty}^\infty dt\,e^{i\omega t}S_{\rm W}(E,t)=S(E+\tfrac{1}{2}\omega,E-\tfrac{1}{2}\omega).\label{SWdefomega}
\end{equation}
In terms of $S_{\rm W}(E,\omega)$ the expression \eqref{Itaverage1} for the current reads
\begin{align}
I(t)={}&\tfrac{1}{2}e\int_{-\infty}^\infty\frac{d\bar{E}}{2\pi}\int_{-\infty}^\infty\frac{d\omega'}{2\pi}\int_{-\infty}^\infty\frac{d\omega}{2\pi}\,e^{i\omega t}\nonumber\\
&\times\bigl[f(\bar{E}-\tfrac{1}{2}\omega')-f(\bar{E}+\tfrac{1}{2}\omega')\bigr]\nonumber\\
&\times {\rm Tr}\,S_{\rm W}^\dagger(\bar{E}+\tfrac{1}{2}\omega,\omega+\omega')\sigma_z S_{\rm W}(\bar{E},\omega'),\label{ItaverageW}
\end{align}
with the definitions $\bar{E}=\frac{1}{2}(E+E')$, $\omega'=E-E'$.

The integrals over $\omega$ and $\omega'$ contribute over the range $-\gamma_0\lesssim\omega,\omega'\lesssim\gamma_0$. To leading order in $\gamma_0$ we  therefore have
\begin{align}
&{\rm Tr}\,S_{\rm W}^\dagger(\bar{E}+\tfrac{1}{2}\omega,\omega+\omega')\sigma_z S_{\rm W}(\bar{E},\omega')=\nonumber\\
&\qquad{\rm Tr}\,S_{\rm F}^\dagger(\bar{E},\omega+\omega')\sigma_z S_{\rm F}(\bar{E},\omega')+{\cal O}(\gamma_0/E_{\rm c}),\label{Trleadingorder}
\end{align}
in view of Eq.\ \eqref{SWSFexpansion}. Substitution into Eq.\ \eqref{ItaverageW}, with a change of variables $\omega''=\omega+\omega'$, results in
\begin{align}
I(t)={}&\tfrac{1}{2}e\int_{-\infty}^\infty\frac{d\bar{E}}{2\pi}\int_{-\infty}^\infty\frac{d\omega'}{2\pi}\int_{-\infty}^\infty\frac{d\omega''}{2\pi}\,e^{i(\omega''-\omega' )t}\nonumber\\
&\times\bigl[f(\bar{E}-\tfrac{1}{2}\omega')-f(\bar{E}+\tfrac{1}{2}\omega')\bigr]\nonumber\\
&\times {\rm Tr}\,S_{\rm F}^\dagger(\bar{E},\omega'')\sigma_z S_{\rm F}(\bar{E},\omega')[1+{\cal O}(\gamma_0/E_{\rm c})]\nonumber\\
={}&\tfrac{1}{2}e\int_{-\infty}^\infty\frac{d\bar{E}}{2\pi}\int_{-\infty}^\infty\frac{d\omega}{2\pi}\,e^{-i\omega t}\nonumber\\
&\times\bigl[f(\bar{E}-\tfrac{1}{2}\omega)-f(\bar{E}+\tfrac{1}{2}\omega)\bigr]\nonumber\\
&\times {\rm Tr}\,S_{\rm F}^\dagger(\bar{E},t)\sigma_z S_{\rm F}(\bar{E},\omega)[1+{\cal O}(\gamma_0/E_{\rm c})].\label{Itavleadingorder}
\end{align}

Since we do not wish to assume that $\gamma_0$ is small compared to $kT$, we expand the difference of Fermi functions in square brackets to all order in $\omega$, 
\begin{align}
&[f(\bar{E}-\tfrac{1}{2}\omega)-f(\bar{E}+\tfrac{1}{2}\omega)]e^{-i\omega t}=\nonumber\\
&\qquad=-2\sum_{p=0}^{\infty}\frac{(\omega/2)^{2p+1}}{(2p+1)!}\frac{\partial^{2p}}{\partial\bar{E}^{2p}}f'(\bar{E})e^{-i\omega t}\nonumber\\
&\qquad=-\left(\sum_{p=0}^{\infty}\frac{(i/2)^{2p}}{(2p+1)!}\frac{\partial^{2p}}{\partial\bar{E}^{2p}}\frac{\partial^{2p}}{\partial t^{2p}}\right)f'(\bar{E})\omega e^{-i\omega t}.\nonumber\\
\end{align}
Upon partial integration, the sum over $p$ contributes to the integral \eqref{Itavleadingorder} terms of order
\begin{equation}
\frac{\partial^{2p}}{\partial\bar{E}^{2p}}\frac{\partial^{2p}}{\partial t^{2p}}\,S_{\rm F}(\bar{E},t)={\cal O}(\gamma_0/E_{\rm c})^{2p},\label{sumoverp}
\end{equation}
so only the $p=0$ term needs to be retained to leading order. 

We thus arrive at
\begin{align}
I(t)={}&-\tfrac{1}{2}e\int_{-\infty}^\infty\frac{d\bar{E}}{2\pi}\int_{-\infty}^\infty\frac{d\omega}{2\pi}\,f'(\bar{E})\omega e^{-i\omega t}\nonumber\\
&\times {\rm Tr}\,S_{\rm F}^\dagger(\bar{E},t)\sigma_z S_{\rm F}(\bar{E},\omega)[1+{\cal O}(\gamma_0/E_{\rm c})]\nonumber\\
={}&-\tfrac{1}{2}ie\int_{-\infty}^\infty\frac{d\bar{E}}{2\pi}\,f'(\bar{E})\nonumber\\
&\times {\rm Tr}\,S_{\rm F}^\dagger(\bar{E},t)\sigma_z \frac{\partial}{\partial t}S_{\rm F}(\bar{E},t)[1+{\cal O}(\gamma_0/E_{\rm c})].
\label{Itavfinal}
\end{align}
At zero temperature, when $-f'(E)\rightarrow\delta(E)$, we recover Eq.\ \eqref{IBPT},
\begin{equation}
I_{\rm adiabatic}(t)=\frac{ie}{4\pi}\,{\rm Tr}\,S_{\rm F}^\dagger(0,t)\sigma_z \frac{\partial}{\partial t}S_{\rm F}(0,t).\label{IBPTapp}
\end{equation}

\section{Multi-channel probe}
\label{app_multi}

\subsection{Coupling matrix}

In the main text we assumed that the pair of Andreev levels near the level crossing is coupled to a \textit{single} pair of electron-hole modes in the normal-metal probe. This coupling is described by the $2\times 2$ coupling matrix $W$ defined in Eq.\ \eqref{HWgeneral}. More generally, a multi-channel probe has a $2\times 2N$ coupling matrix of the form
\begin{equation}
W=(W_1,W_2,\ldots W_N),\;\;W_n=\begin{pmatrix}
\alpha_n&\beta_n^\ast\\
\beta_n&\alpha_n^\ast
\end{pmatrix},\label{Wmulti}
\end{equation}
constrained by particle-hole symmetry: $W=\sigma_x W^\ast\sigma_x$. We collect the complex coefficients $\alpha_n,\beta_n$ in a pair of vectors,
\begin{equation}
\bm{\alpha}=(\alpha_1,\alpha_2,\ldots\alpha_N),\;\; \bm{\beta}=(\beta_1,\beta_2,\ldots\beta_N),\label{alphabetadef}
\end{equation}
and define the inner products
\begin{equation}
\begin{split}
&\langle\bm{\alpha}|\bm{\alpha}\rangle=\sum_{n=1}^N|\alpha_n|^2,\;\;\langle\bm{\beta}|\bm{\beta}\rangle=\sum_{n=1}^N|\beta_n|^2,\\
&\langle\bm{\alpha}|\bm{\beta}\rangle=\sum_{n=1}^N\alpha_n^\ast\beta_n.
\end{split}
\label{innerproducts}
\end{equation}

The decay rates $\gamma_1$, $\gamma_2$ of the pair of quasibound Andreev levels are the eigenvalues of the $2\times 2$ matrix
\begin{align}
WW^\dagger&=\sum_{n=1}^N W_n^{\vphantom{\dagger}} W_n^\dagger\nonumber\\
&=\begin{pmatrix}
\langle\bm{\alpha}|\bm{\alpha}\rangle+\langle\bm{\beta}|\bm{\beta}\rangle&2\langle\bm{\alpha}|\bm{\beta}\rangle^\ast\\
2\langle\bm{\alpha}|\bm{\beta}\rangle&\langle\bm{\alpha}|\bm{\alpha}\rangle+\langle\bm{\beta}|\bm{\beta}\rangle
\end{pmatrix},\label{WWdagger}\\
&\Rightarrow\begin{cases}
\gamma_1=\langle\bm{\alpha}|\bm{\alpha}\rangle+\langle\bm{\beta}|\bm{\beta}\rangle+2|\langle\bm{\alpha}|\bm{\beta}\rangle|,\\
\gamma_2=\langle\bm{\alpha}|\bm{\alpha}\rangle+\langle\bm{\beta}|\bm{\beta}\rangle-2|\langle\bm{\alpha}|\bm{\beta}\rangle|.\\
\end{cases}\label{gamma12WWdagger}
\end{align}
As before, we define the arithmetic and geometric averages,
\begin{equation}
\bar{\gamma}=\tfrac{1}{2}(\gamma_1 +\gamma_2),\;\;\tilde{\gamma}=\sqrt{\gamma_1\gamma_2}.\label{gammabartildedef}
\end{equation}
For later use, we also note that
\begin{align}
W\sigma_z W^\dagger=\sum_{n=1}^N W_n^{\vphantom{\dagger}}\sigma_z W_n^\dagger=\bigl(\langle\bm{\alpha}|\bm{\alpha}\rangle-\langle\bm{\beta}|\bm{\beta}\rangle\bigr)\sigma_z.\label{WsigmazWdagger}
\end{align}

\subsection{Scattering matrix}

Carrying through the same steps as in the single-channel case, we have the following expression for the $2N\times 2N$ scattering matrix $S$ in terms of the $2\times 2$ Green's function $G$:
\begin{align}
&S(E,E')=2\pi\delta(E-E')-2\pi i W^\dagger G(E,E')W,\nonumber\\
&\left(i \gamma_0^2\sigma_z\frac{\partial}{\partial E}+E +i\pi WW^\dagger \right)G(E,E')=2\pi\delta(E-E').\label{GdefMmmulti}
\end{align}
The solution for $G$ has the factorized form \eqref{Gfactorized}, in terms of the $2\times 2$ matrix
\begin{equation}
X(E)=\begin{pmatrix}
u(E)&v^\ast(-E)\\
v(E)&u^\ast(-E)
\end{pmatrix}\label{Xuvdefmulti}
\end{equation}
that solves the homogeneous equation
\begin{equation}
\left(i \gamma_0^2\sigma_z\frac{\partial}{\partial E}+E +i\pi WW^\dagger \right)X(E)=0.\label{Xdefmulti}
\end{equation}
The functions $u$ and $v$ are determined by
\begin{align}
&\gamma_0^2 u'' + (\varepsilon^2+\delta^2-i)u=0,\;\;
\zeta v=i\varepsilon u-\gamma_0 u',
\label{eq:second-order-eq-for-u_multi}\\
&\varepsilon=(E+i\pi\bar{\gamma})/\gamma_0,\;\;\zeta=(2\pi/\gamma_0)\langle\bm{\alpha}|\bm{\beta}\rangle^\ast,\\
&\delta=|\zeta|=\tfrac{1}{2}(\pi/\gamma_0)(\gamma_1 -\gamma_2).
\label{epsdeltadefmulti}
\end{align}
The solution is
\begin{equation}
\begin{split}
&u(E) = e^{i\varepsilon^2/2}\,U( -\tfrac{1}{4}i \delta^2,\tfrac{1}{2}; -i\varepsilon^2  ), \\
&\zeta v(E) = -\tfrac{1}{2}\delta^2 e^{i\pi/4} \,e^{i\varepsilon^2/2}\,U( \tfrac{1}{2}-\tfrac{1}{4}i \delta^2,\tfrac{1}{2}; -i\varepsilon^2  ).
\end{split}
\label{eq:solution-in-terms-of-u_multi}
\end{equation}
Finally, the scattering matrix has the dyadic form
\begin{align}
&S_{nm}(E,E')|_{E>E'}=-\psi_n(E)\psi^\ast_m(-E'),\label{Suvmulti}\\
&\psi(E)=(2\pi/\gamma_0)e^{\pi\delta^2/8}
W^\dagger
\begin{pmatrix}
u(E)\\
v(E)
\end{pmatrix}.\label{psidefmulti}
\end{align}

\subsection{Transferred charge}

Because the scattering matrix is still of rank-one, a single quasiparticle is transferred as a result of the fermion-parity switch, irrespective of the number of channels $N$ in the metal probe. The charge expectation value of this quasiparticle is given by
\begin{align}
Q&=e\int_0^\infty\frac{dE}{2\pi}\psi^\ast(E)\sigma_z\psi(E)\nonumber\\
&=\frac{2\pi e}{\gamma_0^2}\,e^{\pi\delta^2/4}\int_{0}^{\infty}dE\,{u^\ast(E)\choose v^\ast(E)} W\sigma_z W^\dagger
{u(E)\choose v(E)}\nonumber\\
&=\frac{2\pi e}{\gamma_0^2}\,e^{\pi\delta^2/4}(\langle\bm{\alpha}|\bm{\alpha}\rangle-\langle\bm{\beta}|\bm{\beta}\rangle)\nonumber\\
&\qquad\qquad\times\int_{0}^{\infty}dE\,\bigl(|u(E)|^2-|v(E)|^2\bigr).\label{Qintegralmulti}
\end{align}

Comparison with Eq.\ \eqref{Qintegral} shows that the transferred charge for a multi-channel contact differs from that in the single-channel case by a reduction factor
\begin{align}
{\cal R}&=\frac{\langle\bm{\alpha}|\bm{\alpha}\rangle-\langle\bm{\beta}|\bm{\beta}\rangle}{\tilde{\gamma}}\nonumber\\
&=\frac{\langle\bm{\alpha}|\bm{\alpha}\rangle-\langle\bm{\beta}|\bm{\beta}\rangle}{\sqrt{(\langle\bm{\alpha}|\bm{\alpha}\rangle+\langle\bm{\beta}|\bm{\beta}\rangle)^2-4|\langle\bm{\alpha}|\bm{\beta}\rangle|^2}}\in[0,1],\label{Rresult}
\end{align}
independent of the rapidity $\gamma_0$ of the fermion-parity switch.

As a check, we can directly compute the transferred charge in the adiabatic limit from Eq.\ \eqref{IBPT}. Substitution of the frozen scattering matrix at the Fermi level,
\begin{equation}
S_0=1+2\pi i W^\dagger(E_0\sigma_z-i\pi WW^\dagger)^{-1}W,\label{S0defmulti}
\end{equation}
gives the charge
\begin{align}
&Q_{\rm adiabatic}=\frac{ie}{4\pi}\int_{-\infty}^{\infty} dE_0\,{\rm Tr}\,S_{0}^\dagger\sigma_z \frac{\partial S_0}{\partial E_0}\nonumber\\
&\quad=\frac{e}{2}\int_{-\infty}^{\infty} dE_0\,{\rm Tr}\,(E_0\sigma_z+i\pi WW^\dagger)^{-1}W\sigma_z W^\dagger\nonumber\\
&\qquad\qquad\qquad\cdot (E_0\sigma_z-i\pi WW^\dagger)^{-1}\sigma_z\nonumber\\
&\quad=e\bigl(\langle\bm{\alpha}|\bm{\alpha}\rangle-\langle\bm{\beta}|\bm{\beta}\rangle\bigr)\int_{-\infty}^{\infty} dE_0\,
(E_0^2+\pi^2\tilde{\gamma}^2)^{-1}\nonumber\\
&\quad=e{\cal R}.\label{IBPTmulti}
\end{align}

\subsection{Relation of the reduction factor to the Andreev conductance}

The charge reduction factor ${\cal R}$ from Eq.\ \eqref{Rresult} is a property of the coupling matrix of the normal-metal probe to the Josephson junction. It can be expressed in terms of an independently measurable quantity, the Andreev conductance. 

When the normal-metal probe is biased at a voltage $V$, a current $I$ is driven into the grounded superconductor by the process of Andreev reflection. The Andreev conductance $G_{\rm A}=\lim_{V\rightarrow 0}dI/dV$ is related to the scattering matrix $S_0$ at the Fermi level by
\begin{equation}
G_{\rm A}=\frac{e^2}{2h}\,{\rm Tr}\,(1-S_0\sigma_zS_0^\dagger\sigma_z).\label{GAdef}
\end{equation}

Near the level crossing a resonant peak appears in $G_{\rm A}$ as a function of $E_0$, with the Lorentzian line shape
\begin{equation}
G_{\rm A}=\frac{4e^2}{h}\frac{\pi^2\tilde{\gamma}^2}{E_0^2+\pi^2\tilde{\gamma}^2}\bigl(1-{\cal R}^2\bigr).\label{GAE0}
\end{equation}
The resonant peak height of $(4e^2/h)(1-{\cal R}^2)$ directly determines the charge reduction factor.


\begin{thebibliography}{99}
\bibitem{note1} Since $X$ solves a homogeneous equation, the solution is only determined up to a multiplicative constant. This has no effect on the Green's function, because both $X$ and $X^{-1}$ appear in Eq.\ \eqref{Gfactorized}.
\bibitem{note2} To obtain $v$ from $u$ we used the identity $(d/dz)U(a,b;z)=-az^{-b}U(1+a-b,1-b;z)$.
\bibitem{note3} Eq.\ \eqref{PsiX} describes two equivalent copies of the single-particle excitation $\psi(E)$, one at $E>0$ and one at $E<0$. This double-counting is inherent in the Nambu representation of superconducting quasiparticles. It plays no role in Eqs.\ \eqref{Itdef} and \eqref{Qintegral}, where we restrict ourselves to $E>0$. In Eq.\ \eqref{IBPT} we correct for double-counting by replacing the usual prefactor $1/2\pi$ by $1/4\pi$.
\end{thebibliography}
\end{document}